\documentclass[letterpaper,twocolumn,10pt]{article}
\usepackage{usenix}
\usepackage{amsmath}
\usepackage{mathtools}
\usepackage{mathrsfs,amsmath}  
\usepackage[nolist,nohyperlinks]{acronym}
\usepackage{siunitx}
\usepackage{xspace}
\usepackage{graphicx}
\usepackage{amssymb}
\usepackage{subcaption}

\usepackage[normalem]{ulem}
\usepackage{hyperref}
\usepackage[nameinlink]{cleveref}
\usepackage{multirow}
\usepackage{booktabs}
\usepackage{makecell}
\usepackage{tikz}

\usepackage{fancyhdr}
\include{references.bib}

\newcommand{\name}[0]{\texttt{TRICK}\xspace}
\usepackage{tikz}
\usetikzlibrary{shapes.geometric}
\DeclareRobustCommand*\circled[1]{\tikz[baseline=(char.base)]{
            \node[shape=circle,draw,inner sep=1.2pt] (char) {#1};}}

\DeclareRobustCommand*\ov[1]{\tikz[baseline=(char.base)]{
            \node[shape=ellipse,draw,minimum width=2.5em,   
          minimum height=0.8 em,  
          inner sep=0.8pt] (char) {#1};}}

\begin{document}
%
\title{\Large \bf \name: Time and Range Integrity ChecK using Low Earth Orbiting Satellite for Securing GNSS}




\author{Arslan Mumtaz, Mridula Singh \\ CISPA Helmholtz Center for Information Security, Germany \\ arslan.mumtaz@cispa.de, singh@cispa.de}


\maketitle

\begin{abstract}
Global Navigation Satellite Systems (GNSS) provide Positioning, Navigation, and Timing (PNT) information to over 4 billion devices worldwide. Despite its pervasive use in safety critical and high precision applications, GNSS remains vulnerable to spoofing attacks. Cryptographic enhancements, such as the use of TESLA protocol in Galileo, to provide navigation message authentication do not mitigate time of arrival manipulations.  In this paper, we propose \name, a primitive for secure positioning that closes this gap by introducing a fundamentally new approach that only requires two way communications with a single reference node along with multiple broadcast signals. Unlike classical Verifiable Multilateration (VM), which requires establishing two way communication with each reference nodes, our solution relies on only two measurements with a trusted Low Earth Orbiting (LEO) satellite and combines broadcast navigation signals. We rigorously prove that combining the LEO satellite based two way range measurements and multiple one way ranges such as from broadcast signals of GNSS into ellipsoidal constraint restores the same guarantees as offered by VM whilst using minimal infrastructure and message exchanges. Through detailed analysis, we show that our approach reliably detects spoofing attempts while adding negligible computation overhead.
\end{abstract}

\section{Introduction}
Global Navigation Satellite System (GNSS), such as GPS, Galileo, GLONASS, and BeiDou, are widely used for Positioning, Navigation, and Timing (PNT) information \cite{euspa2024_marketreport}. To prevent against GNSS spoofing, various detection schemes at the user equipment (UE) have been proposed \cite{humphreys2013detection,radovs2024recent,schmidt2016survey}, however, these approaches assume a weak adversary and can be compromised by a more capable, resourceful attacker, highlighting the need for defenses that address stronger threat models~\cite{Executive_Order_13905,Need_for_reselinet_PNT_India,Need_for_reselinet_PNT_Europe}.

Cryptographic enhancements, such as the Timed Efficient Stream Loss-Tolerant Authentication (TESLA) protocol \cite{perrig2003tesla} in Galileo \cite{european_gsc_osnma}, made to prevent spoofing attacks on broadcast positioning signals are proven insecure. These authentication schemes constrain an attacker from forging satellite positioning data. However, it does not prevent selective delay attacks \cite{motallebighomi2023location} that exploit clock inaccuracies of the UE for undetectable spoofing attacks.  The underlying vulnerability arises since the victim UEs cannot securely estimate the contribution of the bias in its clock from the measured pseudoranges.  Due to incorrect clock bias estimation, the attacker spoof a victim's UE into calculating an arbitrary false position, even when achieving that spoofed position requires reducing apparent distances with a subset of GNSS satellites whilst increasing with the remaining. 
On the contrary, if the UE could have secure global time, any adversarial manipulation to achieve an arbitrary position spoofing would be infeasible. However, secure and precise clock synchronization requires two way time transfer and accurate apriori knowledge of relative distance between synchronizing nodes \cite{narula2018requirements}, resulting in an inter-dependency problem. 

\vspace{3pt}
The aforementioned problem of secure positioning can be solved by using Verifiable Multilateration (VM) technique \cite{capkun2006secure}. In this technique, two way ranging via Distance Bounding protocol \cite{brands1993distance} is performed with multiple trusted reference nodes. While Distance Bounding prevents distance reduction attack, countering  distance enlargement requires establishing geometric bound using trusted verifier nodes. Any adversary attempting to enlarge the distance of victim UE from one verifier node must reduce it towards atleast one other verifier. It is important to note that a user can only validate the integrity of its position measurements if it lies within the VM defined geometric bound. 
\begin{figure*}[t]
    \centering
    \includegraphics[width=0.6\linewidth]{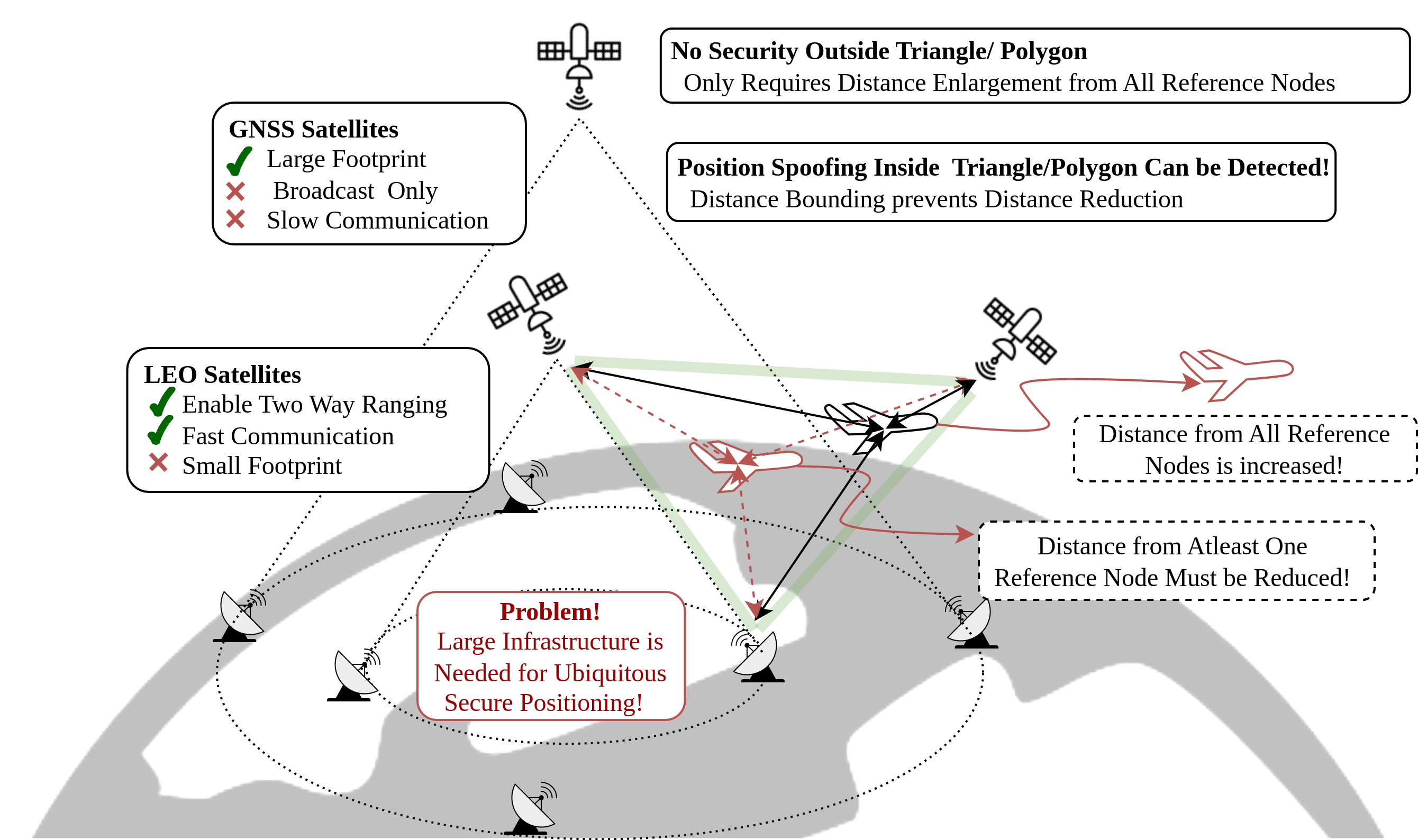}
    \caption{Major problems solved by our proposed mechanisms in \name. }
    
    \label{fig:system_diagram}
\end{figure*}

Realizing VM with Medium Earth Orbit (MEO) Satellites could provide wide area coverage for secure positioning with minimal infrastructure, however, large propagation delays make two way ranging with billions of UEs infeasible. Low Earth Orbiting (LEO) satellites offer a promising alternative \cite{esa_2024_kicks_off_two_new_navigation_missions} due to the shorter propagation delays making them attractive to perform two way ranging via Distance Bounding, however, they have a much smaller coverage area, so achieving a global ubiquitous secure PNT system will require deployment of extremely large constellation.  This requirement of large infrastructure deployment renders VM techniques in their current form impractical in real-world scenarios and demands a secure positioning system that is scalable and integrable. \autoref{fig:system_diagram} illustrates the aforementioned problem. 

\vspace{3pt}
In this paper, we present \name, a secure positioning mechanism that combines two Distance Bounding (DB) measurements (via two way ranging (TWR)) with at least one trusted LEO satellite and passive one-way pseudorange measurements from multiple GNSS satellites. 

While Galileo’s OSNMA, which became fully operational in July 2025, relies on loose time synchronization achieved through two-way communication with a GNSS-independent clock (GIC) source such as NTP or PTP, the proposed \name method utilizes a single Low Earth Orbit (LEO) satellite for two-way communication. Therefore, our proposed   method simply uses the LEO satellite as a reference for GIC instead of terrestrial NTP or PTP server, yet, enables secure position and time estimation within a verifiable geometric bound, without the need for additional infrastructure.  




The core of the proposed mechanism in \name is the use of the UE’s own cryptographically signed challenge transmission time as the single reference for all subsequent time of flight (ToF) calculations. In \name, we use the distance associated to the sum of ToF between the UE and the LEO satellite along with the ToF of GNSS signals, that together define an ellipsoid. Furthermore, we show that any protocol which treats the LEO satellite’s two-way ToF measurement and each GNSS satellite’s one-way pseudorange as independent measurements for multilateration is inherently vulnerable. Specifically, by introducing asymmetric delays in the uplink challenge or the downlink response in the TWR measurement with LEO satellite, an attacker can bias the UE’s clock reference and reduce the apparent GNSS pseudoranges if measured individually, enabling undetected position spoofing. The secure geometric bound formed in our proposed technique is much larger due to the wider coverage area offered by GNSS satellites in MEO, thereby, offering the optimal design in terms of required infrastructure, scalability and security. In summery, our paper makes the following contributions:
\begin{itemize}
    \item Despite decades of work on securing GNSS, it remains vulnerable to selective-delay spoofing. \name closes this gap by introducing a fundamentally new approach of using the sum of the distance for ellipsoidal multilateration to prevent position and time spoofing attacks.
    
    \item Unlike classical verifiable multilateration (VM), which requires multiple bi-directional communications and dense infrastructure, our approach leverages authenticated one-way broadcasts from GNSS satellites to construct secure geometric regions. As a result, \name is the first to integrate distance bounding with broadcast authentication for a scalable and secure positioning system.
    
    \item We rigorously prove that combining the LEO round-trip measurements and GNSS one-way pseudoranges into ellipsoidal constraint restores the same guarantees as offered by VM. 
    
    \item Through detailed analysis of asymmetric delay scenarios, we show that our approach reliably detects spoofing attempts while adding negligible computation overhead.
    
\end{itemize}
\section{Background}
This section provides an overview of how Tesla-based authentication (Galileo OSNMA~\cite{european_gsc_osnma} and GPS CHIMERA~\cite{chimera}) protects GNSS signals from forgery and discuss two way time transfer for clock synchronization~\cite{nist_twoway_transfer}.

\vspace{3pt}
\noindent
\textbf{Galileo OSNMA Signals}

\noindent
\textit{TESLA-Based Authentication:}
Galileo’s OSNMA protocol uses a variant of the TESLA broadcast authentication scheme to secure the E1-B signal’s I/NAV message without affecting legacy Open Service performance. TESLA authenticates streamed data by appending a MAC computed with a secret key from a one-way hash chain. UEs buffer the message until the corresponding key is disclosed, after which the MAC is verified and the data is accepted only if valid.
\vspace{3pt}
\\
\textit{Authentication Delay:}
OSNMA configures TESLA with an Authentication Data and Key Delay (ADKD), typically disclosing the key for a subframe at least 30 seconds later. Each 30-second subframe carries up to six MAC tags, which can only be authenticated after the subframe completes and the corresponding key is received. 
\vspace{3pt}
\\
\textbf{Time Synchronization}
OSNMA enabled Galileo system requires a GNSS-independent clock as the trusted reference node to carry out two-way time transfer and accurately estimate the clock bias using PTP or NTP. In two way time transfer, UE timestamps an outgoing request, the trusted reference returns its current time along with a timestamp of receipt, and the UE finally marks the time of the response according to its local clock. Assuming zero processing delay (for the sake of simplicity), from these three timestamps the UE can conclude that the trusted reference node’s clock leads or lags its own by some \(\delta\) time offset using the following formula:
\begin{equation}
    \delta = \frac{(t_2 - t_1) - (t_3- t_2)}{2}
\end{equation}
In the above equation, \(t_2\) represents the timestamps of the reference node, whereas \(t_1\) and \(t_3\) are the timestamps of the UE according to its local clock. As long as clock drift between these exchanges remains negligible, or is compensated by periodic resynchronization, this loose bound is sufficient to prevent an adversary from forging MAC keys ahead of schedule.
\vspace{3pt}
\\
\textbf{Selective Delay Attacks on GNSS} 
The intrinsic late key disclosure of TESLA creates a vulnerability window. Because the UE does not verify any packet until after the key is revealed, an adversary controlling the signal path can selectively delay each of the GNSS signals without detection \cite{motallebighomi2023location}. Once the UE finally obtains the corresponding key and performs the MAC check, it will accept the delayed packets as valid, even though they arrived 10-100 milliseconds delayed due to malicious activity by an attacker. 
\vspace{3pt}
\\
\textbf{Physical Layer Attacks } 
GNSS signals are also vulnerable to early-detect/late-commit (ED/LC) attacks, where an adversary predicts part of the signal and transmits early to appear closer \cite{clulow2006so}. For encrypted PRN codes with 10 MHz chipping rates \cite{scott2021_gps_galileo}, this can shift perceived arrival time by up to 100 ns which is about 30 meters \cite{zhang2019effects}.  Furthermore, to defend against distance reduction caused by ED/LC and to enable secure two-way ranging with LEO satellites, Leo-Range design \cite{coppola2025leo-range} adopts an OFDM-based signal design. Accordingly, in this paper, we restrict our analysis to attacks that attempt to reduce the estimated distance by selectively delaying GNSS signals, without considering ED/LC vulnerabilities addressed by Leo-Range.

\section{System Design} 
In this section, we begin by briefly defining our threat model and outlining the requirements for a secure PNT system, explained by a VM based mechanism that independently performs Distance Bounding with multiple verifier nodes. From these system requirements, we propose a framework that provides secure positioning, navigation, and timing guarantees equivalent to VM, yet requires minimal bidirectional exchanges, augmented by multiple GNSS broadcast signals.
\subsection{Threat Model}
We focus on a scenario where a UE is interested in securely estimating its position and synchronizing its clock with the global time in the presence of an adversary.
\vspace{3pt}
\\
The UE may or may not have its last known position coordinates as well as ephemeris data \cite{nasa2025_broadcast_ephemeris} of GNSS or any other trusted reference nodes in its local memory i.e, the UE may initiate the procedures of secure positioning after a cold start. However, we assume that the UE must be able to communicate with a trusted reference node, such as LEO satellite for two way ranging and can receive the authenticated broadcast positioning signals. 
\vspace{3pt}
\\
Since, the goal of the adversary \(\mathcal{A}\) is to perform spoofing without being detected, it must avoid jamming that may lead to denial of service and therefore its detection. In order to achieve this goal, we consider a Dolev-Yao's adversary \(\mathcal{A}\) \cite{dolev1983security}, that may position itself anywhere on the line-of-sight paths between the UE and each reference node (the LEO or any GNSS satellite). Under such adversary model, an attacker can intercept, delay, relay and inject signals on both uplink (i.e, from UE to LEO satellite) and downlink (from LEO satellite to the UE) as well as signals arriving from the GNSS satellites.  All positioning signals, both the UE’s two-way challenge/response exchange with the LEO and each GNSS satellite’s  broadcast signals are protected by cryptographic primitives. Consequently, adversary \(\mathcal{A}\) can only learn each positioning signal once it is transmitted by the trusted entities and cannot forge the content of these signals. Any attempt to modify the data bits used as the challenge/response or the cryptographically authenticated data of the broadcast positioning signals will be detected by the UE. 
Although cryptographic defenses at the physical layer bound ED/LC attacks aimed at distance reduction attacks~\cite{coppola2025leo-range, UWB-PR}, an adversary can still reduce the measured range by selectively delaying authenticated signals in GNSS. Because selective delay attacks bypass authentication mechanisms, it is stealthy and poses a greater practical threat. Accordingly, we focus our analysis on selective delay distance reduction attacks (with the standard assumption that no information travels faster than light).

\subsection{Objective system} \label{obj_system}
\begin{figure}
    \centering
    \includegraphics[width=0.7\linewidth]{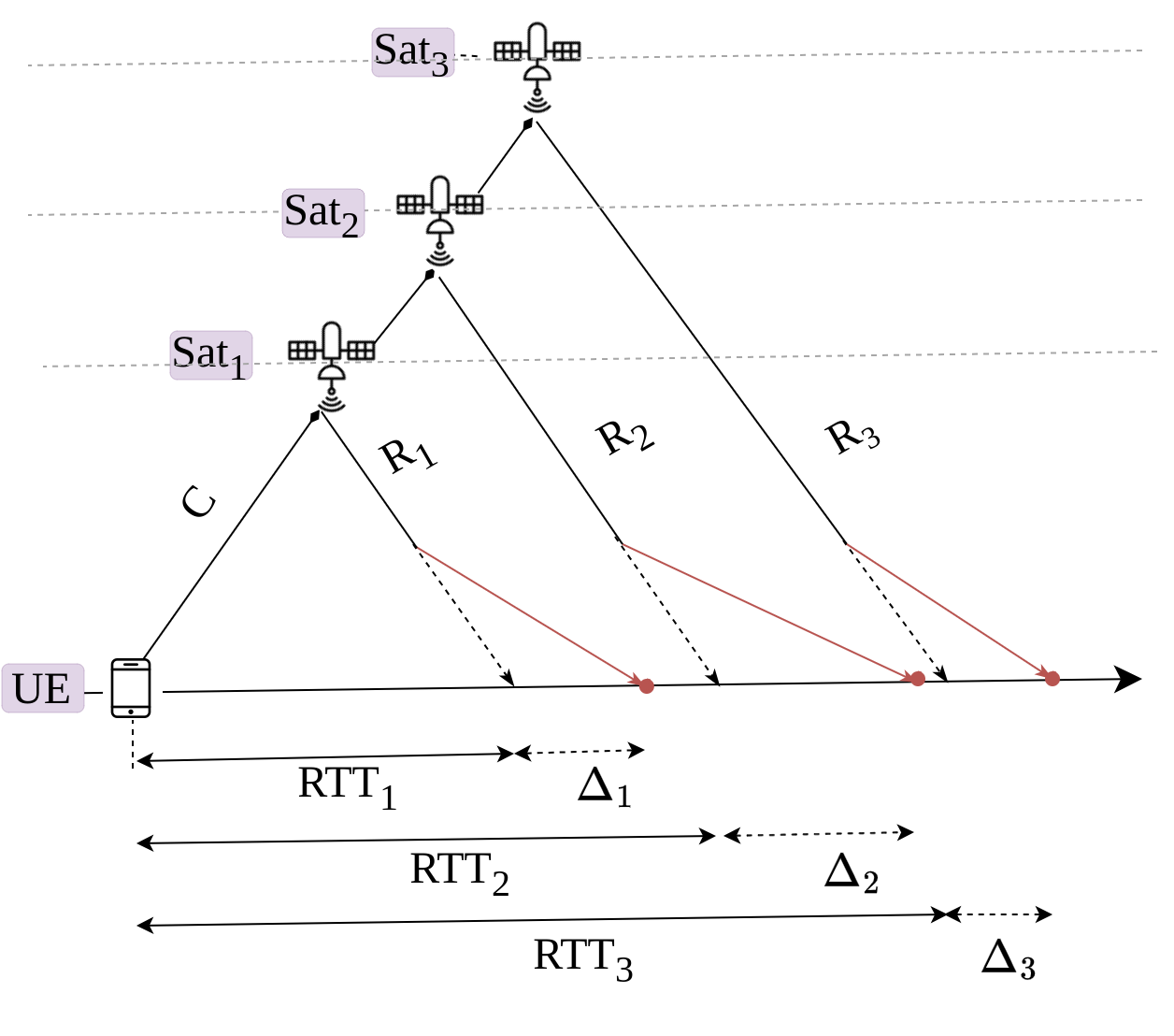}
    \caption{Protocol for performing VM. The described scenario assumes two way communication between UE and each of the LEO satellites.}
    \label{fig:verifiable_multilateration}
\end{figure}

Securely determining a user’s position can be done by distance bounding with several trusted verifier nodes, whose known positions form a VM region.  As shown in \autoref{fig:verifiable_multilateration}, the UE sends the same authenticated challenge nonce (broadcast) to each verifier and immediately starts its own clock to measure round trip times. Using one shared nonce for all the verifier nodes instead of transmitting different one is equally secure and more efficient. Upon receipt of the nonce, each verifier applies cryptographic authentication, and returns the message with its minimal processing delay. The UE then derives an upper bound on its distance to each verifier from the round trip time measurements of the challenge/response exchanges, and the intersection of these bounds yields a precise position estimate.
However, a man in the middle adversary can introduce delay attacks, increasing all RTT measurements and thus enlarging the computed distances to every verifier. Any spoofing attempt inside the polygon formed by verifiers can be detected.
Note that if the measured position lies outside the polygon defined by verifier nodes, the UE cannot distinguish whether this is due to geometry or an attack. Distinguishing between the two requires the UE to already know that its true position lies within that polygon, which may necessitate making the polygon very large.
From the security properties of VM, we derive the following requirements for a secure, scalable, and ubiquitous PNT system: (1) Given the high deployment cost of VM needing simultaneous visibility of at least three LEO satellites everywhere on the Earth, the proposed design should be scalable, use minimal infrastructure, and minimize message exchanges. (2) All verifier nodes require accurate clocks and must communicate the processing delays to the UE. Attackers must be unable to decrease measured ranges by only introducing delays. (3) The verifier‐node polygon must be sufficiently large for the UE to be confident that it lies inside it.
\subsection{\name} \label{steps}
In the following, we present our proposed mechanism in details. First, we provide a concise overview and outline our assumptions and then, we describe the steps required to achieve secure positioning.   
\vspace{3pt}
\\
\textbf{Overview and Assumptions:} To address the challenges of secure and scalable PNT, \name\ combines authenticated broadcasts with at least two secure distance bounding measurements. These two measurements may be obtained either (i) simultaneously from two reference nodes, or (ii) from a single reference node at two different time  instants separated by a fixed duration T (if the reference node moves at high speeds, such as LEO satellites). With authenticated broadcast GNSS signals, \name enables security guarantees equivalent to VM. 
\vspace{5pt}\noindent
\\
Without the loss of generality, in \name, LEO satellite may serve as the reference node for Distance Bounding with the UE, while authenticated GNSS signals can be used to create a scalable, secure PNT system.  We require that every positioning signal, both the LEO challenge–response and the GNSS broadcasts are cryptographically authenticated to prevent forging of node positions and their transmission times i.e., their freshness can be verified. We further assume that cryptographic keys are pre‐shared between the UE and the trusted LEO satellite. Unlike VM, we assume that all the reference nodes, LEO satellites, have globally synchronized clock.  Under these assumptions, we first describe the steps for a single distance bounding measurement and develop the ellipse-based primitive by combining it with broadcast GNSS measurements. We then detail how a second distance bounding measurement, obtained spatially using a second LEO satellite or temporally using the same reference LEO satellite, is incorporated to achieve the full security guarantees of \name.
\subsubsection{Secure Primitive of \name:}
\begin{figure}[t]
    \centering
    \includegraphics[width=0.65\linewidth]{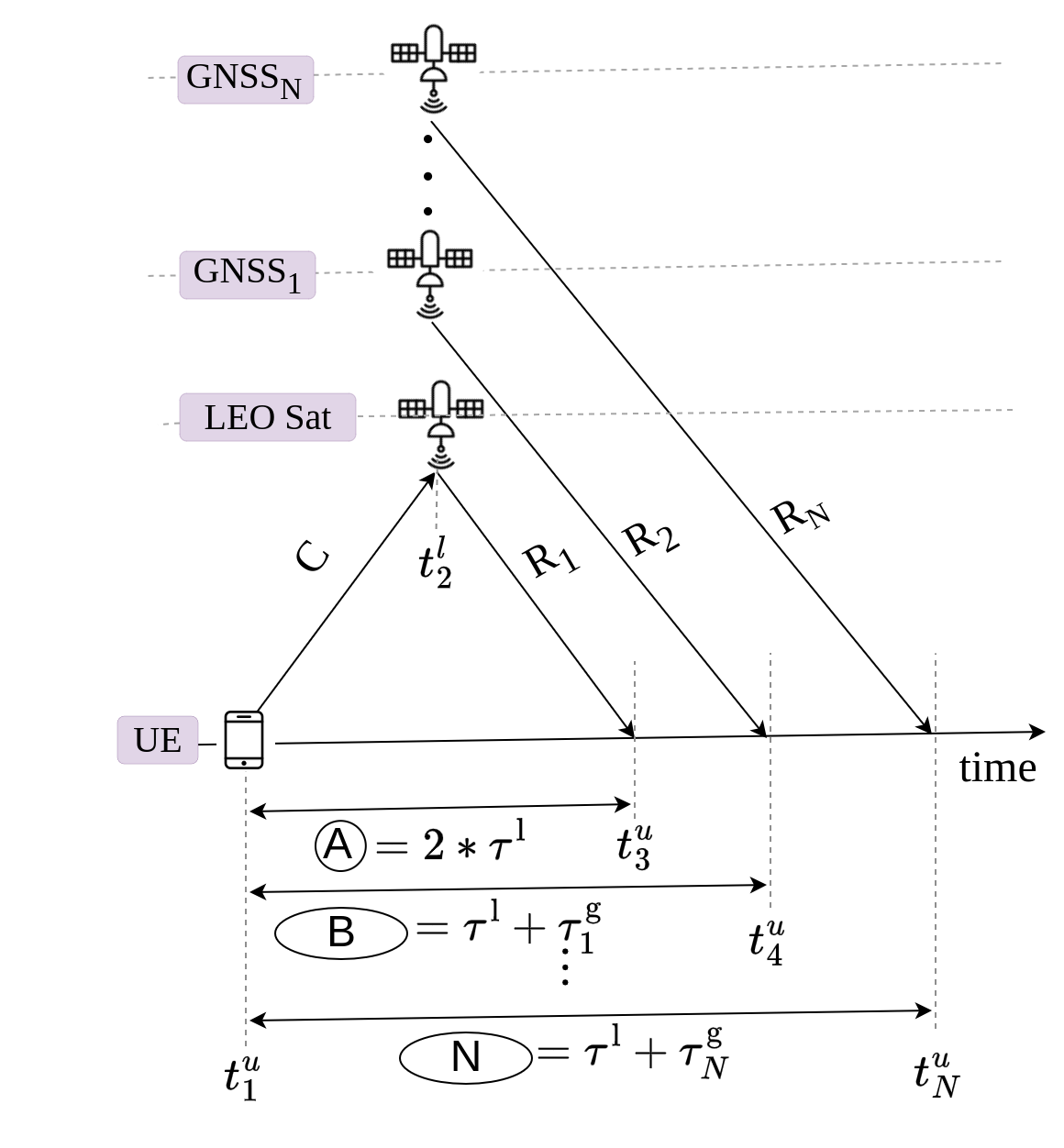}
    \caption{The proposed protocol according to the mechanisms of \name to build ellipse based primitive needed for secure positioning.}
    \label{fig:single_trick_mechanism}
\end{figure}

In the following, we describe the steps that are required to obtain the measurements needed to build an ellipse based \name primitive that is fundamental to provide secure positioning in our design.
\vspace{5pt}\noindent
\\
\textbf{\circled{1} UE Uplink Transmission}
 To obtain secure positioning, the UE generates a cryptographically authenticated nonce and transmits it to the reference LEO satellite as the challenge signal to perform Distance Bounding as shown in \autoref{fig:single_trick_mechanism}. Simultaneously, upon the uplink transmission, it records the exact local transmission time and begins measuring the elapsed interval for subsequent measurements of time of arrival of signals from LEO and the GNSS satellites. 
\vspace{10pt}\noindent
\\
\noindent
\textbf{\circled{2} LEOs Downlink Transmission } Upon reception, the LEO satellite records the precise time of arrival of the UE’s challenge signal. It then verifies the nonce, and computes the response message. Consequently, the response must have protection to enable cryptographic authentication of the challenge's arrival time, the transmission time of the response, processing delay and the satellite's position coordinates along with other necessary information. It must be highlighted that in Distance Bounding protocol, or its use in VM mechanism, a trusted reference is only required to communicate elapsed processing delay. In contrast, \name also requires timestamping of the transmission time for the response message according to the globally synchronized clock.  
\vspace{5pt}\noindent
\\
\textbf{\circled{3} ToA Estimation at the UE } The UE receives the downlink of the reference LEO satellite and estimates its precise time of arrival. Following the reception, the UE cryptographically verifies the response signal to ensure the integrity of the encrypted timing and ephemeris information. The UE can then compute the RTT \circled{A} between itself and the LEO satellite using this arrival time, the processing delay of the LEO satellite and the previously marked reference transmission time \(t^u\) according to its local clock as depicted in \autoref{fig:single_trick_mechanism}. The UE subsequently receives the GNSS signals (OSNMA and CHIMERA protocol) that are cryptographically protected utilizing the TESLA protocol and estimates their time of arrival. We compute the duration between each of the arrival time of the GNSS signal and the transmission time of the challenge signal transmitted to LEO satellite from its local clock. As shown in \autoref{fig:single_trick_mechanism}, this duration represents the combined ToF from UE to LEO \(\tau_{\text{u}}^{\text{l}}\) plus the ToF from GNSS satellite to the UE \(\tau_{\text{i}}^{\text{g}}\). Rather than computing separate ranges of UE to each GNSS satellite from their time difference of arrival, \name directly uses the sum \(\tau_{\text{u}}^{\text{l}}+\tau_{\text{i}}^{\text{g}}\) represented by \ov{B} in \autoref{fig:single_trick_mechanism} for first GNSS, so on to \ov{N} for the nth GNSS satellite used in the position computation. In this step, we require only the combined ToF measurements on UE's local clock in order to provide an upper bound on the sum of the distances to each LEO–GNSS pair. An attacker cannot reduce the total accumulated propagation time, therefore, each measurement provides an upper bound similar to the conventional Distance Bounding protocol. In the security analysis \autoref{sec_analysis}, we will describe in details how using the sum of the distance can detect malicious attacks caused by introducing asymmetric delays.
\begin{figure}[t]
\centering
\includegraphics[width=0.7\linewidth]{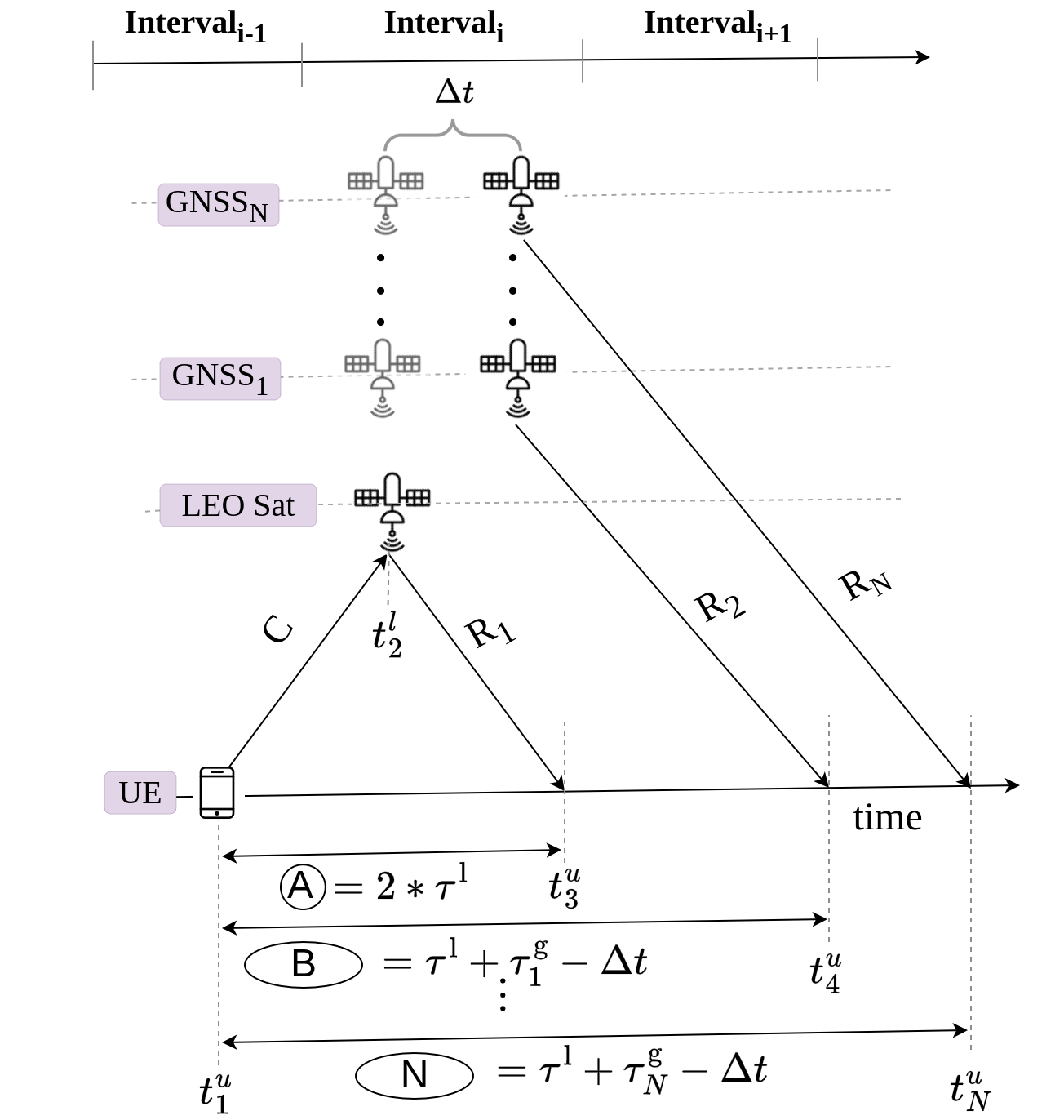}
\caption{ A Scenario where the downlink transmission of LEO satellite is not synchronized with the GNSS satellites.}        
\label{fig:non_synched_downlink}
\end{figure}
\vspace{5pt}\noindent
\\
\textbf{\circled{4} Synchronized Downlink Transmission } We require that downlink transmission of the response signal is synchronized with the transmission of the encrypted GNSS broadcast signals as shown in \autoref{fig:single_trick_mechanism}. However, \name compensates systems with unsynchronized downlink transmissions by requiring precise time synchronization across all reference nodes and the inclusion of their timestamps in the downlink positioning signals. The security implications justifying the rationale of this assumption are examined in detail in Appendix A. 
The UE must introduce an offset term \(c \cdot \Delta t_i\) into equation for the sum of all distance measurements described in previous step. This offset term represents the difference between the LEO response’s transmission time (extracted from the response signal) and the GNSS broadcast time (cryptographically authenticated) as shown in \autoref{fig:non_synched_downlink}, and is defined as:
\begin{equation}
    \Delta t_i = t^l -t^g_i
\end{equation}
 Where \(t^l\) is the transmission time of the LEOs response signal and \(t^g_i\) is the transmission time of the ith GNSS satellite. Even though GNSS satellites send synchronized downlinks, \name works without such synchronization among broadcast nodes, as long as the positioning signal includes both its transmission timestamp and the associated ephemeris data. By explicitly modeling \(c \cdot \Delta t_i\) for each measurement, it can create synchronized framework compatible even with all asynchronous downlink transmitting reference nodes. The UE must carefully incorporate the offset term \(\Delta t_i\) based on the following considerations:
\begin{itemize}
    \item \(\Delta t_i < 0\) is the scenario when the GNSS transmission follows the transmission of the LEOs' response as shown in \autoref{fig:non_synched_downlink}.  In that case the UE receives the LEO’s authenticated timestamp before the GNSS signals, and simply adds a negative offset to obtain the essence of synchronized downlink transmissions.         
    \item \(\Delta t_i > 0\) is the scenario when the GNSS transmission precedes the LEO satellite's response. The UE must add a positive offset, in effect pushing the GNSS timestamp forward, so that all measurements can be made equivalent to synchronized transmissions. However, this scenario is critical for OSNMA’s TESLA based authentication, where delayed key disclosure duration is less than \(\Delta t_i\). This is because if the UE accepts the GNSS broadcast signals that are transmitted earlier than the downlink transmission of the LEO response signal by more than the delayed key disclosure duration, an attacker could exploit this to launch GNSS forgery attacks. 
\end{itemize}
\textbf{\circled{5} Position Computation }  The UE formulates the equations of ellipsoid by treating the LEO and GNSS satellite positions as the foci to determine its own coordinates and the sum of the distance being the constant from the reference nodes as discussed in previous step. The following equation formulates the relation as:
\begin{equation} \label{eq:ellipsoid}
   || r^g_{i} - r^{u} || + || r^{l} - r^{u} || = c. [\tau_{\text{u}}^{\text{l}}+\tau_{\text{i}}^{\text{g}} +  (t^l -t^g_i)] 
\end{equation}
where \(r^g_{i}\) and \(r^l\)  are vectors of known position coordinates of GNSS and LEO satellites respectively, \(r^{u}\) is the vector of unknown position coordinates for the UE and c is the speed of the light. 
The UE then performs ellipsoidal multilateration by solving the equation of ellipsoids for its three unknown position coordinates using a numerical solver.
\vspace{5pt}\noindent
\\
\textbf{\circled{6} GNSS Data Integrity Check } \name assumes that the data integrity of the GNSS signals remains intact. If an adversary can forge the signals without being detected, the security guarantees cannot be provided. The OSNMA enabled Galileo signals preserve integrity by having each UE loosely synchronize its local clock to a global reference time. To achieve this, the UE and an external clock must perform two way time transfer. As mentioned earlier, secure clock synchronization requires not only two way time transfer, but precise knowledge of the distance between the reference node and the UE resulting in an interdependency that can undermine security if not verified.
\vspace{5pt}\noindent
\\
\textbf{\textit{Loose Time Synchronization}} \name addresses this challenge by assuming that each UE knows the maximum possible ToF from any LEO satellite given the knowledge of orbits. This assumption is verifiable because a UE can only receive a signal from a LEO satellite only if that is above the minimum elevation angle required for communication. The UE uses this bound as part of its integrity check during loose clock synchronization, prior to calculating a secure position and time according to mechanisms of \name. This knowledge bounds the maximum spoofing in the clock bias that an attacker can introduce at the UE by delaying the signal's arrival time.  If known, the UE compares the maximum possible ToF with the RTT measurement computed using Distance Bounding protocol to detect any malicious delay introduced by an adversary.

It’s important to note that \name’s integrity check is only applicable when the duration in TESLA protocol’s delayed key disclosure exceeds the maximum possible ToF between the UE and the LEO satellite in a given constellation. In Galileo’s OSNMA implementation, this late key disclosure delay is 30 seconds, while the worst case LEO ToF is on the order of tens of milliseconds. Therefore, this condition will always be satisfied.  
\begin{figure*}[t] 
    \centering
    \includegraphics[width=1\linewidth]{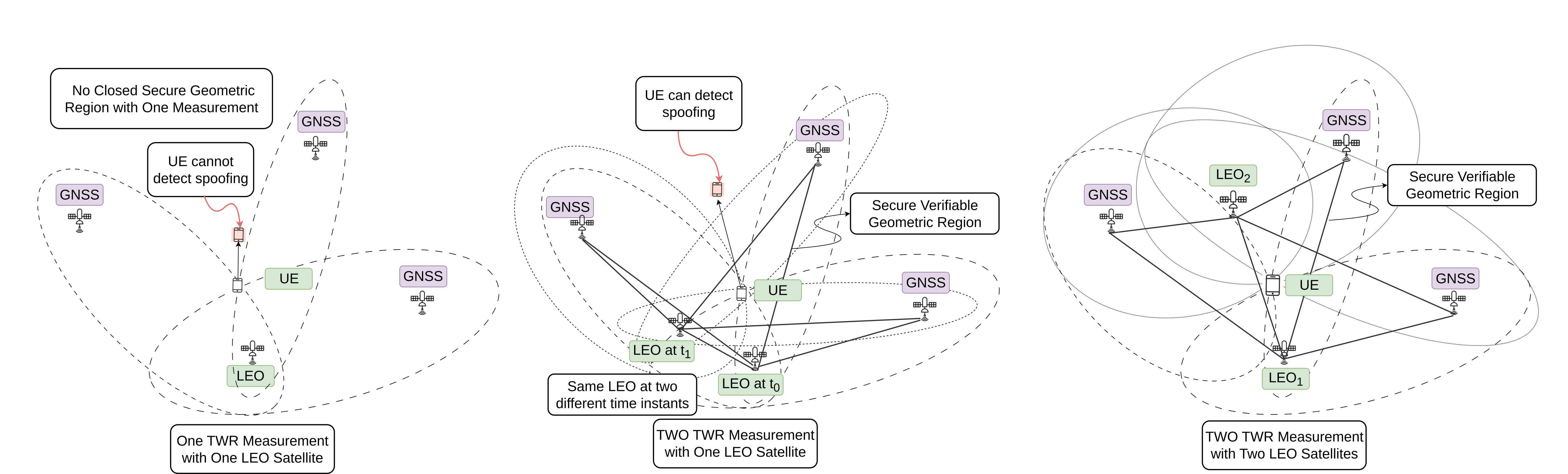}
    \caption{\textbf{Left:} A single TWR and broadcast measurement forms ellipses that only increase. However, UE cannot detect spoofing due to lack of closed regions. \textbf{Middle:} Two time separated measurements with a moving LEO form secure triangles with GNSS, enabling spoofing detection. \textbf{Right:} Instant spoofing detection is possible via measurements with two LEO satellites. }
    \label{fig:Secure_triangles}
\end{figure*}
\subsubsection{Secure Positioning using \name Primitive}
\textbf{Secure Geometric Region:} A single distance bounding measurement provides authenticated upper bound on the combined propagation time from a UE to LEO and GNSS satellite to UE. However, to obtain security guarantees for the computed position by the user, we require a closed geometry around the target area, constructed from the known position coordinates of the anchor nodes (such as LEO satellite and GNSS satellites), so that the UE is provably contained within it and cannot be spoofed anywhere inside such region by merely adding non-negative delays to the sum of propagation time as shown in \autoref{fig:Secure_triangles}. To achieve this, \name utilize two independent distance bounding measurements constraints by having the UE perform two-way ranging with two trusted LEO satellites. Alternatively, the same guarantee is obtained if the UE performs two-way ranging with the same LEO at two different time instants, provided the satellite’s position changes during the chosen interval. In both cases (two LEOs or one LEO at two times), the two distance bounding measurements together with any authenticated GNSS broadcast define a closed triangular acceptance region. If an adversary only increases the measured sums of distances (by adding delay), the UE computed position will be outside this triangle. To spoof the UE inside such triangle, the adversary would have to reduce at least one authenticated sum of distances, which is not feasible according to the primitive of \name (see \autoref{sec_analysis} for proof). Thus, the second distance bounding measurement turns \name bounds into a closed, verifiable region for acceptance.
\vspace{5pt}\noindent
\\
\textbf{First Integrity Check:} After computing its position, the UE runs the first integrity check. For each visible GNSS satellite, it forms a triangle using either (i) the two LEO satellite references used for two-way ranging  or (ii) the same LEO satellite reference at the two measurement instants, together with that GNSS satellite. It must be mentioned that if the UE is on the surface of the Earth, the triangle uses subsatellite points, otherwise a reference node on the Earth surface is required for three dimensional positioning. The estimated position of the UE is accepted only if it lies inside at least one such triangle, which realizes the closed acceptance region described above with multiple GNSS satellites.In the Evaluation \autoref{evaluation}, we will perform comparative analysis on how the use of broadcast GNSS nodes having large coverage area and orbiting at higher altitude is helping the user to obtain optimized, secure and ubiquitous PNT service instead of relying on LEO satellites only infrastructure.  
\vspace{5pt}\noindent
\\
\textbf{Second Integrity Check:} The second integrity check verifies that, for each visible GNSS satellite and the reference LEO satellite, the combined geometric distance from the UE’s estimated position (after solving ellipsoidal method) to the LEO plus to that GNSS satellite exactly matches the sum of the corresponding measured pseudoranges sum. In practice, once the UE has computed a position using the ellipsoidal technique, it recomputes the straight line distance to the LEO and to each GNSS satellite from that position. It then adds those two values together and compares the result against the initially measured sum. Any successful spoofing inside the triangular region for all GNSS and LEO reference nodes requires at least one of the measured sum to be reduced below its true value. As a consequence, the sum of measured pseudoranges will mismatch the sum of the recomputed sum of distances after UE solves for its position if an adversary attempted to spoof the position of the victim inside the formed regions. The UE flags this discrepancy by checking whether the difference between computed sum of distances and sum of measured pseudoranges exceeds a small threshold. This threshold value can be selected based on the standard deviation of pseudorange errors caused by channel conditions, as evaluated in Section~\ref{sec:threshold}.

\begin{figure*} 
    \centering
    \includegraphics[width=1\linewidth]{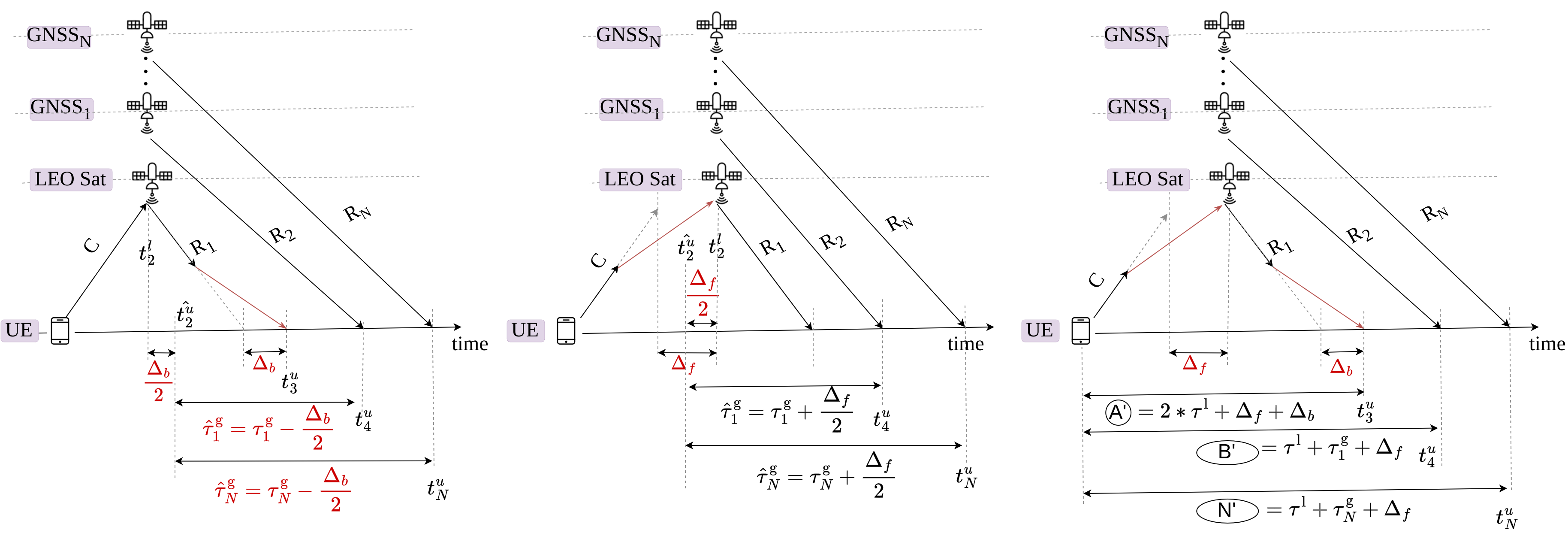}
    \caption{(\textbf{Left:}) The adversary introduces positve delay in the LEO’s downlink signal to bias the UE's clock, thereby increasing the measured two‐way  ToF measured with LEO satellite. After compensating clock bias, all one way GNSS pseudoranges appear reduced by \(\frac{\Delta_b}{2}\) . By further applying per GNSS spoofing offsets, attacker can spoof the victim in to computing any fake target position. (\textbf{Middle:}) The adversary delays only the UE’s uplink to the LEO, causing each distance (LEO and GNSS) to increase after incorrect reference of the local clock bias. Because no one way GNSS pseudorange is reduced, the attacker cannot reduce the distances and cannot spoof a position to arbitrary fake target. But victim cannot differentiate this case from the one where backward delay is greater. (\textbf{Right:}) Since, \name uses sum of ToF with LEO and GNSS, it can detect selective delay attacks that can only cause an increase in the sum of the distances with the appropriate use of the reference. It can be seen that any forward or backward delay cannot reduce the measured sum.}
    \label{fig:security analysis}
\end{figure*}
\section{Security Analysis} \label{sec_analysis}
This section provides the detailed security analysis of our proposed primitive \name used for secure positioning and prove that it can detect position and time spoofing attacks. To this end, we  prove that any protocol which treats the LEO’s two-way ToF and each GNSS satellite’s one-way ToF independently for pseudorange estimation and then applies multilateration is insecure. In specific, by introducing asymmetric delays, in the UE’s challenge signal (forward-leg) or the LEO’s response signal (backward-leg), an adversary can manipulate the time reference for the UE such that the measurements of pseudoranges of the GNSS signals can be reduced in comparison to their correct value. Consequently,  we model both forward delay and backward delay attacks to demonstrate this fundamental vulnerability. 
\subsection{Backward Delay Attack} 
We consider the first attack scenario where the adversary \(\mathcal{A}\) introduces no delay on the ToF of the challenge signal transmitted by the UE in the uplink (i.e, \(\Delta_f = 0\)), but imposes a positive delay on the ToF of the response signal transmitted by the LEO satellite (i.e, \(\Delta_b > 0\)) as shown in the left of \autoref{fig:security analysis}. In general, this scenario captures the concept where backward delay is greater than the forward delay ( \(\Delta_b > \Delta_f\)). In the two‐way exchange with the LEO under attack by \(\mathcal{A}\), the UE obtains three timestamps: \(t_{1}^{u}\), the local transmit time of its challenge; \(t_{2}^{l}\), the reception (and instantaneous retransmission) time at the LEO; and \(t_{3}^{u}\), the time at which the UE receives the response. The instantaneous retransmission is assumed with zero processing delay at the LEO here for the sake of simplicity, but in reality there will be non-negligible positive delay that can be factored out. Representing the true ToF by \(\tau^{l}\) and the true clock offset by \(\delta\), the timestamps satisfy:
\begin{equation} \label{t2 and t3}
\begin{aligned}
t_{2}^{l} &= t_{1}^{u} + \tau^{l} + \delta,\\
t_{3}^{u} &= t_{2}^{l} - \delta + \tau^{l} + \Delta_{b}.
\end{aligned}
\end{equation}
From these observations the UE can derive the ToF from LEO satellite:
\begin{equation} \label{tof leo}
\hat{\tau}^{l}
= \frac{t_{3}^{u} - t_{1}^{u} }{2}
\end{equation}
where \(\hat{\tau}^{l}\) represents the spoofed ToF resulting from the manipulation of the attacker. Since the attacker introduced the positive delay in the backward leg, the total round trip time of the signal according to the local clock of the UE is increased compared to its correct value. Therefore, we can write the relation between the correct ToF and the spoofed ToF as follows:
\begin{equation} \label{spoofed tof leo}
\hat{\tau}^{l} = {\tau}^{l} + \frac{\Delta_{b}}{2}
\end{equation}
Since, the UE is aware that the transmission time of all the GNSS satellites and the LEO satellite is same according to the depiction shown in the \autoref{fig:security analysis}, therefore, UE aims to compute the reference time of transmission of LEO \(t_{2}^{l}\) using the authenticated timestamps such that it can compute the individual ToF associated to all the GNSS satellites separately. This requires the UE to compute the clock bias of its local clock and compute the local reference that must be equal to the transmission of the downlink signals.
The local clock bias at the UE can be computed in the following manner:
\begin{equation} \label{estimated clock bias}
\hat{\delta}
= \frac{(t_{2}^{l} - t_{1}^{u}) - (t_{3}^{u} - t_{2}^{l})}{2}
\end{equation}
 By solving the above equation, we can write the relation between the correct clock bias and the spoofed clock bias in the following way:
 \begin{equation} \label{spoofed_clock_bias}
 \hat{\delta} = \delta - \frac{\Delta_{b}}{2}
\end{equation}
Using the above clock bias, the UE can compute the local clock reference for the transmission time \(t_{2}^{l}\) of the downlink signal as shown in the left of \autoref{fig:security analysis} and given by:
 \begin{equation}
 \hat{t_{2}^{u}} = t_{1}^{u} + \hat{\tau}^{l} - \hat{\delta}
\end{equation}
Expansion of the above equation and solving gives the relation between the UE computed time under adversary influence and the correct value in the following:
\begin{equation}
 \hat{t_{2}^{u}} = t_{2}^{l} +  \frac{\Delta_b}{2} - \delta
\end{equation}
As the UE has compensated for clock bias in the computation of local reference of the time of transmission of the downlink signals by all satellites, therefore, the arrival time of the subsequently GNSS satellite can be written in the following way: 
\begin{equation}
 t_{i}^{u} = t_{2}^{l} +  \hat{\tau_i}^{g} - \delta
\end{equation}
The ToF can then be simply computed as the difference of time of arrival of GNSS signal and the computed transmission time according to the local clock of the UE as shown in the left of \autoref{fig:security analysis}, resulting in:
\begin{equation}
 \hat{\tau_i}^{g} = \tau_i^{g} - \frac{\Delta_b}{2}
\end{equation}
From the above equation and illustration in the \autoref{fig:security analysis}, it is clear that UE computed ToF corresponding to the GNSS satellite is reduced by a factor of \(\frac{\Delta_b}{2}\). Consequently, an attacker can further introduce selective delays in individual GNSS satellites to adjust the distance corresponding to the fake target position, even if a subset of it requires distance reduction. This violates the requirements outlined earlier in \autoref{obj_system} that an adversary must not be able to reduce the distance from any of trusted reference node to provide secure position and time estimate using VM.
\subsection{Forward Delay Attack}
Now consider the other case where the adversary \(\mathcal{A}\) introduces no backward delay (\(\Delta_{b}=0\)) but imposes a positive forward delay \(\Delta_{f}>0\) on the ToF of the transmitted challenge signal from UE to the LEO satellite as shown in the middle of \autoref{fig:security analysis}.  By construction, this cannot reduce the estimated distances to either the LEO or any of the GNSS satellite. To observe it, we model the effect of the forward delay in the following similar to \autoref{t2 and t3} as:
\begin{equation}
t_{2}^{l} = t_{1}^{u} + \tau^{l} + \delta + \Delta_{f}
\end{equation}
 whereas the arrival time of the response signal having zero backward delay can be written as:
\begin{equation}
    t_{3}^{u} = t_{2}^{l} - \delta + \tau^{l}
\end{equation}
From these, the UE can compute the estimated ToF with LEO satellite and its local clock bias same as \autoref{tof leo} and \autoref{estimated clock bias}, which is given as:
\begin{equation}
\hat{\tau}^{l}
= \frac{t_{3}^{u} - t_{1}^{u}}{2}
= \tau^{l} + \frac{\Delta_{f}}{2},
\end{equation}
\begin{equation}
\hat{\delta}
= \frac{(t_{2}^{l} - t_{1}^{u}) - (t_{3}^{u} - t_{2}^{l})}{2}
= \delta + \frac{\Delta_{f}}{2}.
\end{equation}
When the UE uses these to set its local reference transmit time and then measures each GNSS pseudorange, the resulting estimates are:
\begin{equation}
\hat{\tau}_{i}^{g}
= \tau_{i}^{g} + \frac{\Delta_{f}}{2},
\end{equation}
\noindent i.e.\ every inferred distance is strictly larger than the true value as shown in left of \autoref{fig:security analysis}.  Hence, \(\mathcal{A}\) cannot execute any distance‐reduction attack under \(\Delta_{f}>0,\,\Delta_{b}=0\), and VM over these pseudoranges remains secure. However, the UE cannot distinguish this case from the first attack scenario of greater backward delay.  
\subsection{Analysis of \name}
To defend against both attack scenarios, we form a single combined ToF measurement associated to LEO and for each GNSS satellite \(i\) as shown in the right of the \autoref{fig:security analysis} and given as:
\begin{equation}
    t_{i}^{g} - t_{1}^{u}  =\hat\tau^{\,l} + \hat\tau_{i}^{\,g}
\end{equation}
and enforce the ellipsoidal constraint
\begin{equation}
     || r^{l} - r^{u}|| + || r_{i}^{g} - r^{u}|| = c.(t_{i}^{g}- t_{1}^{u})
\end{equation}
 \\
\textbf{Backward Delay Attack (\(\Delta_{f}=0, \Delta_{b}>0\)) }
As derived earlier, from the two way exchange with LEO satellite, in the presence of the adversary, we have associated ToF given as 
\begin{equation}
  \hat\tau^{l} = \tau^{l} + \frac{\Delta_{b}}{2}
  \end{equation}
  Similarly, for the subsequently arriving GNSS signals as derived previously:
  \begin{equation}
       \hat\tau_{i}^{g} = \tau_{i}^{g} - \frac{\Delta_{b}}{2},
  \end{equation}
so that
\begin{equation}
  \hat\Sigma_{i}
  = \bigl(\tau^{\,l} + \tfrac{\Delta_{b}}{2}\bigr)
  + \bigl(\tau_{i}^{\,g} - \tfrac{\Delta_{b}}{2}\bigr)
  = \tau^{\,l} + \tau_{i}^{\,g}.
\end{equation}
Any additional positive delays \(\Delta_{i}^{g}\) on the GNSS link only increase \(\hat\tau_{i}^{g}\) and hence
\(Sum \ge\tau^{\,l}+\tau_{i}^{\,g}\).  Thus the ellipsoid’s constant can only increase,  and cannot be reduced, even in the backward delay scenario. 
\\
\textbf{ Forward Delay Attack (\(\Delta_{b}=0, \Delta_{f}>0\)).}
Similarly, we can also derive for the forward delay scenario in the following way:
\begin{equation}
  \hat\tau^{l} = \tau^{l} + \frac{\Delta_{f}}{2}
\end{equation}
\begin{equation}
    \hat\tau_{i}^{g} = \tau_{i}^{g} + \tfrac{\Delta_{f}}{2}
\end{equation}
so that
\begin{equation}
  Sum
  = \bigl(\tau^{\,l} + \frac{\Delta_{f}}{2}\bigr)
  + \bigl(\tau_{i}^{\,g} + \frac{\Delta_{f}}{2}\bigr)
  = \tau^{l} + \tau_{i}^{g} + \Delta_{f}
\end{equation}
which again can only increase.
In both cases, any adversarial delay causes the ellipsoid constant sum to be \(\ge\tau^{l}+\tau_{i}^{g}\), so the computed ellipsoid grows or remains the same.  Consequently, an attacker can no longer perform a distance‐reduction attack, and we recover the same provable security guarantees as VM.
\section{Secure Clock Synchronization}
After performing the two integrity checks in \autoref{steps}, the UE can trust its position estimate, but this does not gaurentee the sure clock bias estimate.
\vspace{5pt}\noindent
\\
\textbf{\textit{Clock Bias Estimate}} An attacker may apply a backward delay to the LEO response without altering GNSS pseudoranges, thereby bypassing both position checks as the backward delay cancels out (\autoref{sec_analysis}). To detect this, the UE verifies that the measured RTT to the LEO matches the geometric 2*ToF derived from its verified position. Any mismatch reveals manipulation of time of arrival of the Response signal received from LEO, indicating clock spoofing.
\vspace{5pt}\noindent
\\
\textbf{Clock Drift Estimate}
\name records multiple clock bias estimates over time to compute the drift rate (i.e., the rate at which bias evolves). This measured slope is compared against a pre-characterized bound of the local oscillator’s drift \cite{anderson2025time}. If the observed drift exceeds the bound, an attack is detected. For adversaries attempting signal time-stretching or compression, we refer to dedicated detection methods in \cite{anliker2023time}, which is outside the scope of this paper.
\section{Evaluations} \label{evaluation}
\textbf{\textit{Attack on GIC based GNSS Positioning:}} We begin our evaluation of \name by analyzing a spoofing attack on conventional GNSS positioning systems that rely on a GNSS-independent clock (GIC) for time synchronization. Specifically, we first demonstrate how manipulating this GIC reference allows adversaries to spoof multilateration-based GNSS positioning. We then evaluate \name under the same threat model and show that it successfully detects such spoofing attempts.
\vspace{5pt}
\\
\textbf{GIC manipulation:}
Galileo OSNMA receivers require a trusted GIC to achieve synchronization. If an adversary compromises this step—e.g., by attacking the reference node providing clock synchronization—the UE’s clock becomes biased, allowing spoofing of all GNSS pseudoranges. We demonstrate this by assuming a LEO satellite acts as the reference node (without loss of generality) and show that manipulating its downlink can shift the UE’s computed position to any desired fake location, even one that would require reducing some GNSS pseudoranges.
\vspace{5pt}\noindent
\\
\textbf{Error in Range:} To simulate real-world GNSS errors, we model pseudorange noise using standard techniques (similar to GPS-SDR-SIM \cite{GPS_SDR_SIM}), incorporating thermal noise, multipath, atmospheric effects, and hardware bias. Each pseudorange equals the true geometric range plus independent noise, matching the expected error distribution of civilian-grade receivers (i.e., tens of nanoseconds, or a few meters).
\vspace{5pt}\noindent
\\
\textbf{Spoofing Signal Generators:} Typical GNSS spoofers \cite{GPS_SDR_SIM} generate signals with configurable pseudorange offsets, but cannot manipulate the GIC reference. To fully emulate asymmetric delay attacks, we inject a positive delay into the LEO downlink (used for clock synchronization) in addition to generating noisy GNSS pseudoranges. This allows us to evaluate the combined attack in a controlled simulation settings. As a result, we emulate all such scenarios in for comprehensive analysis.
\begin{figure}[t]
    \centering
    \includegraphics[width=0.95\linewidth]{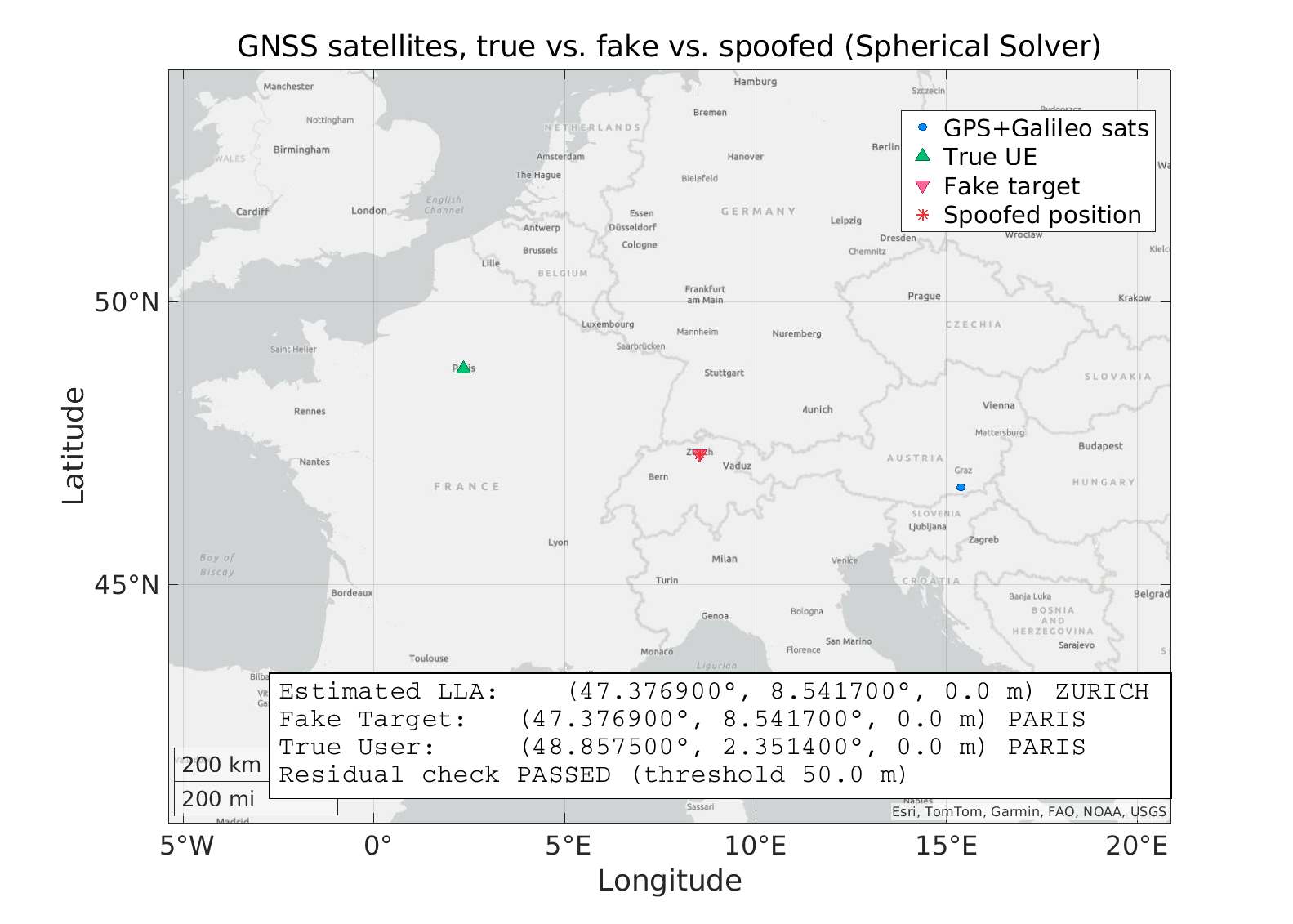}
    \caption{Blue markers indicate the subsatellite locations of all visible GPS and Galileo satellites; The victim’s true positon (Paris) is shown by a green triangle, while the attacker’s chosen fake target and the receiver’s resulting (spoofed) estimate both coincide at Zurich (downward red triangle and red asterisk).}
    \label{fig:attack_geoplot}
\end{figure}
\vspace{5pt}\noindent
\\
\textbf{Attack Steps} The adversary selects a fake target position within the polygon formed by visible GNSS and LEO satellites. A backward delay is applied to the LEO two-way ranging link, shifting the UE’s clock. If the required GNSS delays to spoof the target remain non-negative, the attack is deemed feasible. \autoref{fig:attack_geoplot} illustrates such an attack from Paris to Zurich. A 1 ms backward delay biases the clock, and selective GNSS delays cause the UE to compute the fake position using spherical multilateration.
\begin{figure}[t]
    \centering
    \includegraphics[width=0.9\linewidth]{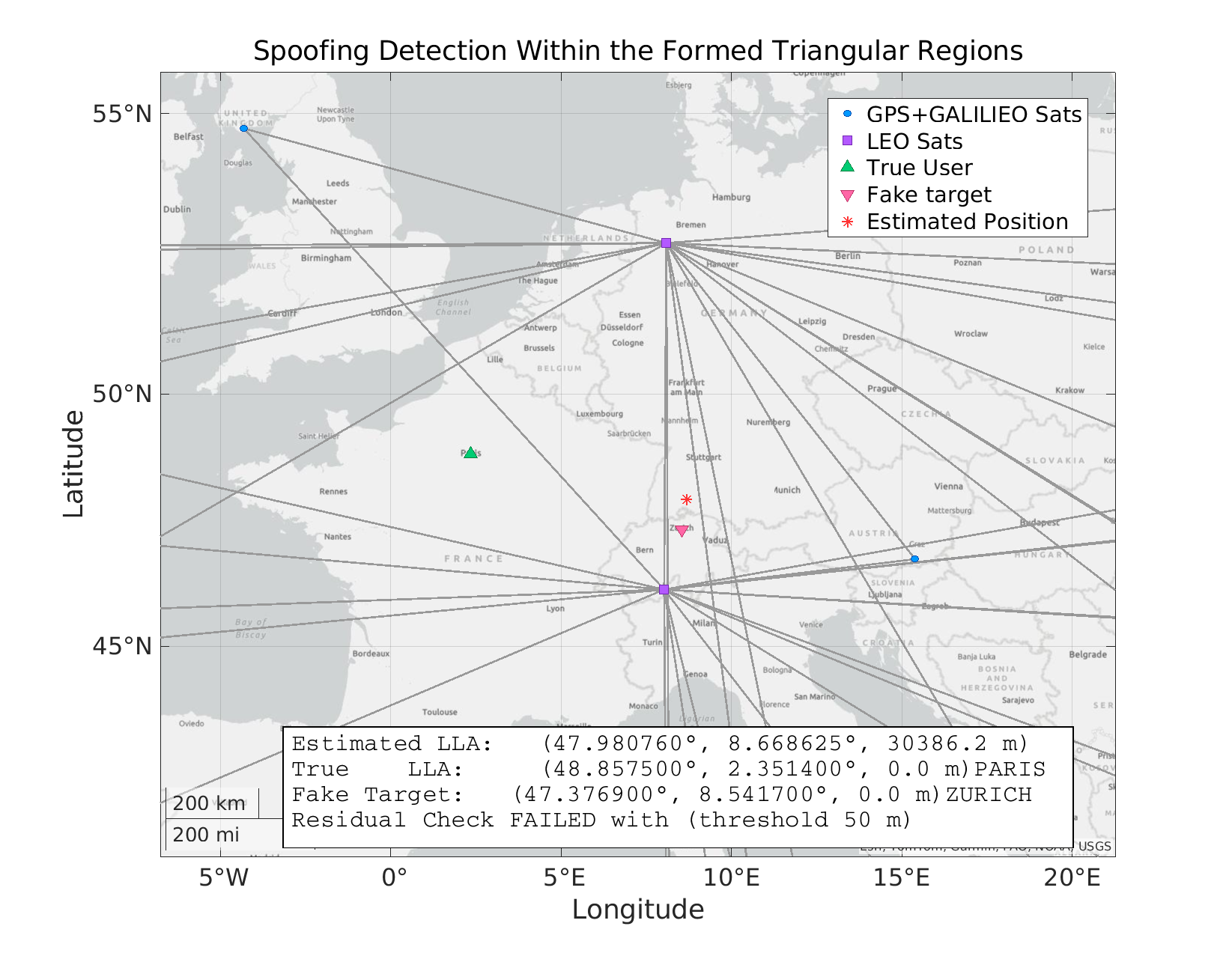}
    \caption{ Using our proposed technique, the red asterisk shows the UE’s computed (spoofed) position that is incorrect by over 30 km from the fake target position and lies inside the formed polygon. Although the false position passes the first integrity check, the second integrity check (residual of measured sum vs. geometric sum) fails, immediately detecting the spoofing attempt.}
    \label{fig:geo_plot_single_trick}
\end{figure}
\subsection{Detection using \name}
\label{sec:threshold}
In our implementation of \name, \autoref{fig:geo_plot_single_trick} shows the detection of the previously described attack. In this scenario, we launch the same attack that we discussed against GNSS receiver. Even though the computed position of the victim using our proposed ellipsoidal method falls inside the polygon using spoofed measurements, however, the second integrity check fails as there is a mismatch in the sum of the distance at the spoofed position and the initial measurements. As a result, the residual error for the measured sum of the distance of LEO satellite and each of the GNSS satellite from the UE computed position exceeds the threshold (we choose a threshold of 50 meters for each sum).
\begin{figure}[t]
    \centering
    \includegraphics[width=1\linewidth]{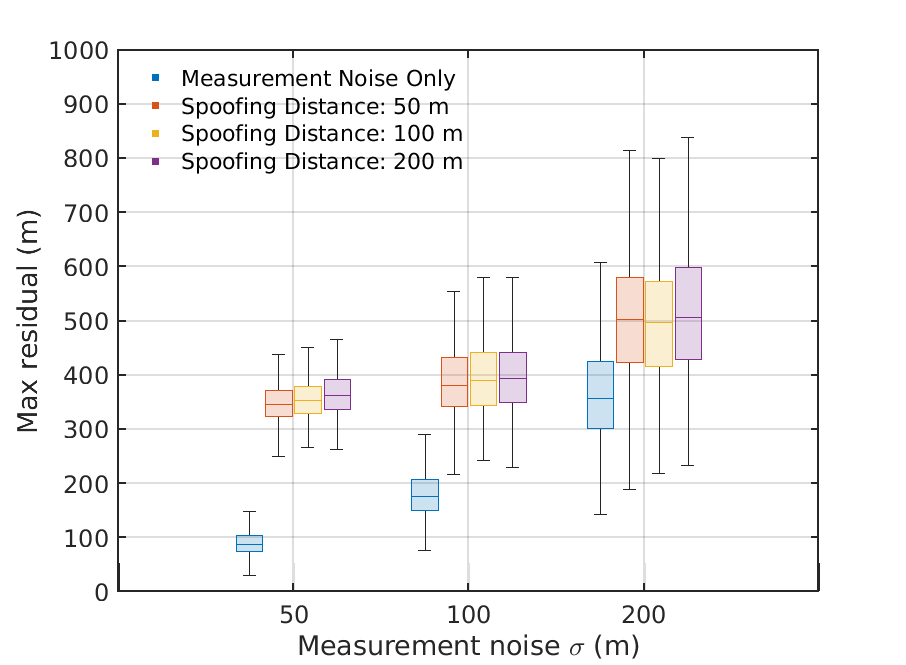}
    \caption{Box plots of maximum residual errors for spoofing distances of 50 m, 100 m, and 200 m under measurement noise levels of 50 m, 100 m, and 200 m, showing how the distribution of residual for second integrity grows with both noise and spoofing range.}
    \label{fig:residual_error_with_ellipsoid_solver}
\end{figure}
\vspace{5pt}\noindent
\\
\textbf{\textit{Distribution of Residual Errors }} To evaluate the robustness of the second integrity check under spoofing attempts and varying noise levels, we simulate the \name protocol for a range of adversarial configurations. Specifically, we simulate spoofing offsets of 25m, 50m, and 100m from the true user position, each evaluated under additive pseudorange noise levels of 50m, 100m, and 200m (standard deviation). In each simulated trial, we compute the user’s position using the ellipsoid solver and evaluate the second integrity check by recomputing the geometric distance sum from the estimated position to both the LEO and each GNSS satellite. We then compute the absolute residual between this predicted sum and the measured pseudorange sum. \autoref{fig:residual_error_with_ellipsoid_solver} visualizes the distribution of these maximum residuals using box plots. Each group on the x-axis corresponds to a different noise level, while the colored boxes represent different spoofing distances. It can be seen that spoofing displacements induce significant residual shifts well beyond the corresponding noise levels. For instance, when the spoofing offset is just 25m and the measurement noise is 50m, the minimum observed residual exceeds 200m. This indicates that the solver reliably detects anomalies that go beyond what would be expected from measurement noise alone. Even at larger spoofing distances (50m and 100m), the resulting residuals stay consistently higher and clearly separated from the noise induced variance. 
When spoofing to a geographically distant location, the residual error becomes significantly more pronounced. For example, in the cases discussed above, spoofing a location in Paris to appear as one in Zurich results in an estimated residual error of approximately 30km.

\begin{figure*}[t] 
    \centering
    \includegraphics[width=1\linewidth]{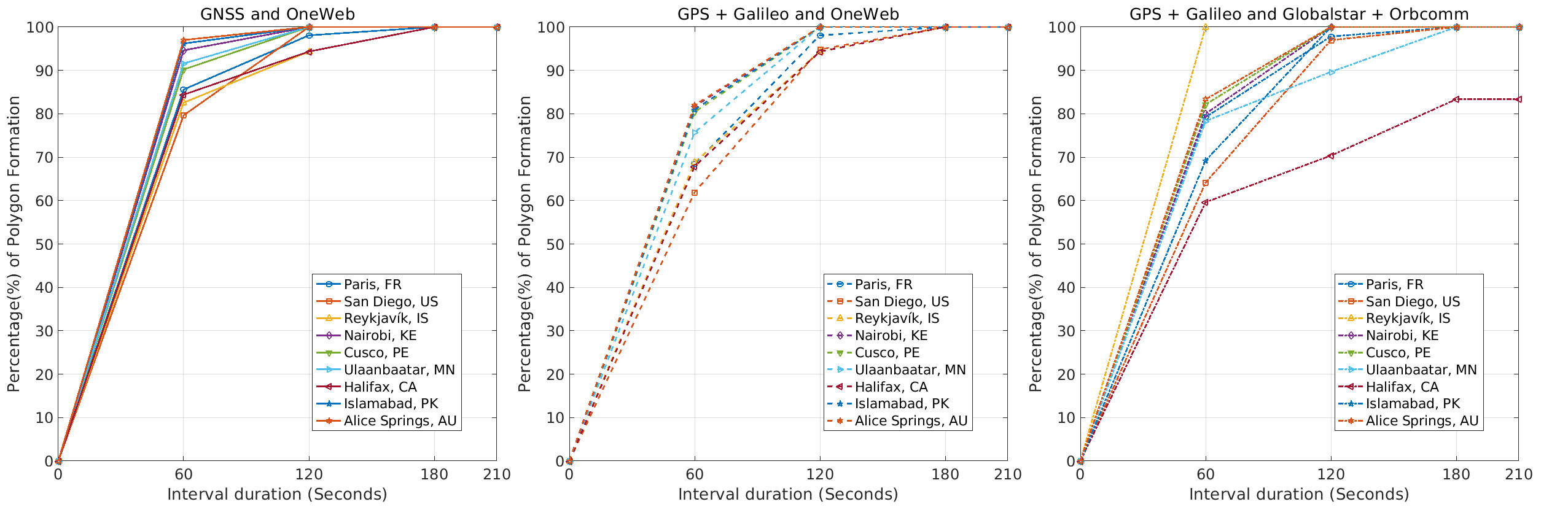}
    \caption{Availability of secure geometric regions formed via two distance bounding exchanges with a moving LEO and GNSS satellites, evaluated across global ground stations for varying time separations. Each plot shows the percentage of time a valid acceptance triangle can be formed under different satellite constellation configurations.}
    \label{fig:sweep}
\end{figure*}
\begin{figure}[t]
    \centering
    \includegraphics[width=1\linewidth]{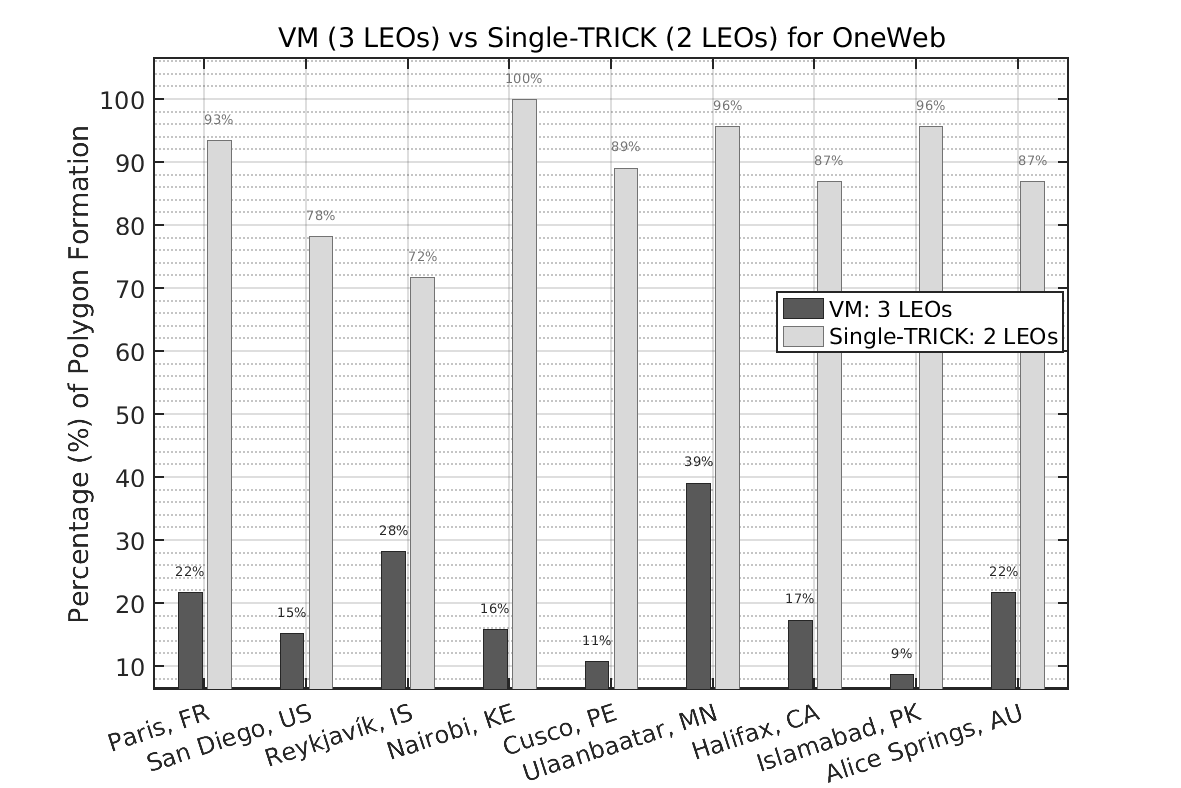}
    \caption{Percentage of Polygon Formations for Diverse Geographical Regions over Orbital Period of Satellites}
    \label{fig:vm_vs_trick}
\end{figure}

\vspace{5pt}\noindent
\\
\textbf{\textit{Coverage of Secure Triangular Regions}}
We investigate the availability of secure triangular acceptance regions, which are needed for the geometric security guarantees in our scheme. Such a region is defined by two distance bounding measurements and GNSS broadcast signals, where these measurements come from either (i) two distinct LEO satellites visible simultaneously or (ii) the same LEO satellite observed at two instants separated by an interval of $\Delta T$. 
\vspace{5pt}\noindent
\\
\textbf{\textit{Results using Single Satellite}}
we evaluate the availability of secure geometric regions when the user performs two TWR measurements with the same LEO satellite, separated by a time interval $\Delta T$. This forms a temporal triangle with a GNSS satellite, allowing a closed acceptance region to be defined. Our objective is to determine the optimal separation interval $\Delta T$ that maximizes the likelihood of triangle formation. \autoref{fig:sweep} shows the percentage of coverage for different intervals. To ensure a comprehensive and geographically diverse analysis, we select nine representative ground stations across the globe. These include locations in Europe (Paris), North America (San Diego and Halifax), South America (Cusco), Africa (Nairobi), Asia (Ulaanbaatar and Islamabad), and Oceania (Alice Springs), as well as a high-latitude site in Iceland. These sites span a wide range of latitudes and regional configurations, ensuring that our visibility and coverage analysis reflects global conditions rather than being biased by regional characteristics. For the LEO infrastructure, we consider moderate sized constellations that are already operational. We simulate over the full orbital period of each considered LEO constellation to capture a complete cycle of satellite visibility and ensure representative coverage statistics across time. Specifically, in the first plot we simulate  OneWeb  as a LEO constellation and GNSS satellites that include various constellations. In the second plot, we analyze the one web with GPS and Galileo constellations only, whereas the third plot involves Orbcomm and Globabstar satellite constellations. At every sampled evaluation time, we determine whether the user is able to form any valid triangle. The triangle is defined using the subsatellite points of the respective anchors, and the user's estimated position must lie within at least one such triangle to be accepted.
\vspace{5pt}\noindent
\\
\textbf{\textit{Comparison of VM and \name}}
In the second configuration, we analyze the availability of secure positioning using \name when the user performs simultaneous distance bounding with two distinct LEO satellites. Each pair of LEO positions and any visible GNSS satellite can form a triangle around the user’s position, enabling the same geometric acceptance region as before. This setting enables us to compare \name directly against VM, which requires three simultaneous LEO satellites for position verification.
We use the OneWeb constellation for this comparison because it provides sufficient LEO visibility for both schemes; in contrast, constellations like Orbcomm and Globalstar have significantly fewer satellites, making it infeasible to consistently form closed triangles using three LEOs. Using the same set of UE location, we simulate both VM and our \name scheme over OneWeb’s orbital cycle and measure the fraction of time each method can form at least one triangle around the UE.
\autoref{fig:vm_vs_trick} summarizes the results in a histogram. Each bar represents the availability of secure triangle formation under each scheme, per location. Across all locations, \name consistently outperforms VM in availability, even though it uses only two LEO references. This highlights the practical benefit of our approach. By combining a pair of LEO measurements with authenticated GNSS broadcasts, \name achieves higher coverage while reducing the infrastructure requirement compared to VM.
\section{Discussion}
\textbf{Duration for Secure Positioning Fix:}
In Galileo OSNMA and GPS CHIMERA, the time to first authenticated fix (TTAF) is typically 2–3 minutes due to the low data rates of their CDMA-based channels and the slow delivery of cryptographic material. The cryptographic material is received in fragmented segments that must be collected and validated over time before a position fix can be obtained. Similarly, if we rely on a single LEO satellite for secure time and position fix, the duration required to form the secure polygon region remains comparable to that of existing GNSS systems like Galileo OSNMA or GPS CHIMERA, ensuring that our design does not introduce any additional delay beyond what is already required by authenticated GNSS positioning. Once this authenticated fix is obtained, secure position estimates can be computed instantly using one-way pseudoranges from GNSS satellites. Alternatively, for continuous secure positioning, the UE may perform periodic distance bounding (e.g., every second) and maintains a sliding window of time tagged secure ranges, which it uses with GNSS pseudoranges to form triangles at each new time instant.
\vspace{5pt}\noindent
\\
\textbf{Integration with 5G-NTN.} \name can be seamlessly integrated into 5G non-terrestrial networks to deliver both secure PNT and low-latency communication at scale. First, the 5G core natively manages subscriber identities and keys, enabling Distance Bounding without additional trust infrastructure. Second, 3GPP-compliant LEO satellites with regenerative payloads act as in-orbit base stations capable of demodulating, processing, and timestamping uplink/downlink signals. This allows direct support for \name’s challenge–response exchanges and secure ToF measurements.
\vspace{5pt}\noindent
\\
\textbf{Distributed Anchors.}
\name assumes a trusted LEO satellite with correct time reference. However, trust in a single anchor poses a risk. LEO satellites may be spoofed or lack security patches due to hardware or other kind of limitations. To address this, \name can be extended to multi-hop Distance Bounding, where time and range integrity is verified across multiple nodes. Full protocol design is left for future work. 
\vspace{5pt}\noindent
\\
\textbf{Ground Infrastructure as Anchor.}
While \name uses LEO satellites as secure anchors due to their GNSS based synchronized clocks, the design is not satellite specific. Any ground node with a trusted, GNSS synchronized clock can serve as an anchor, enabling secure positioning. However, the tradeoff is reduced coverage compared to satellite based configurations.
\section{Related Work}
Several countermeasures leveraging spatial diversity~\cite{Nils2011requirements}, inertial sensors~\cite{GPS_aanjhan,GPS_harshad}, advanced signal processing~\cite{Aanjhan_Spree,Akos2012WhosAO}, opportunistic signals~\cite{HybridPositioning_5G_1,HybridPositioning_5G_2} and bounding clock drift \cite{anderson2025time} have been studied to increase resilience against GNSS spoofing~\cite{Spoofing2,GPS_Spoofing1}. However, these countermeasures are often application-specific and do not prevent physical layer attacks~\cite{gnss-wasp,motallebighomi2023location}. We provide a fundamental design, that is application agnostic, and does not require any additional hardware, such as inertial sensors. 

GNSS systems are using cryptographic solutions to enhance their security guarantees. Galileo introduced OSNMA~\cite{european_gsc_osnma} to authenticate the navigation message contents using TESLA protocol~\cite{perrig2003tesla}. In addition, GPS and Gelilio are introducing hidden markers at the physical layer ~\cite{chimera,european_gsc_osnma}. Since in GNSS, the user’s location is computed based on both the navigation message contents and its time-of-arrival, these systems are still vulnerable to selective delay attacks~\cite{motallebighomi2023location}. While selective delay attacks can be prevented using tight time synchronization with a GNSS independent clock, the clock synchronization protocols are vulnerable to asymmetric delay attacks, when user position is unknown~\cite{narula2018requirements}. In our system we break this dependency, and shows the feasibility of acquiring both time and position simultaneously. Moreover, we leverage security guarantees provided by cryptographic solutions of the GNSS  to make our system scalable.

Since GNSS spoofing attacks affect critical infrastructure, alternative solutions are being explored~\cite{JRC132737}. LEO based PNT system are gaining attention due to the reduced cost of launching LEO satellites~\cite{iannucci2020economical}.  Even though the idea of navigation using LEO satellite is not new, these system are explored to provide opportunistic and dedicated PNT services~\cite{iannucci2020economical,reid2018broadband,orabi2021opportunistic,neinavaie2021blind}. For example, Iridium constellation of LEO satellites is providing time and location services \cite{Satelles}. However, literature lacks the security analysis of these positioning techniques, and high strength of the signals transmitted by LEO satellite is often cited as the security feature against spoofing attacks. 

The two-way ranging systems that use Distance Bounding protocols~\cite{Brands1994,DB_Ariadne_forRouting,DB_AccessControl} to upper bound distance measurement and to be used for VM are mostly designed for the short ranging positioning infrastructure, using Ultrawideband and WiFi based infrastrcture~\cite{W-SPS}. The most recent work~\cite{coppola2025leo-range} (USENIX Security 2025), demonstrate the feasibility of distance bounding using LEO satellites. Nonetheless, their does not exist any study on how these designs should be used for the ubiquitous secure positioning. 

\begin{figure}[t]
    \centering
    \includegraphics[width=0.85\linewidth]{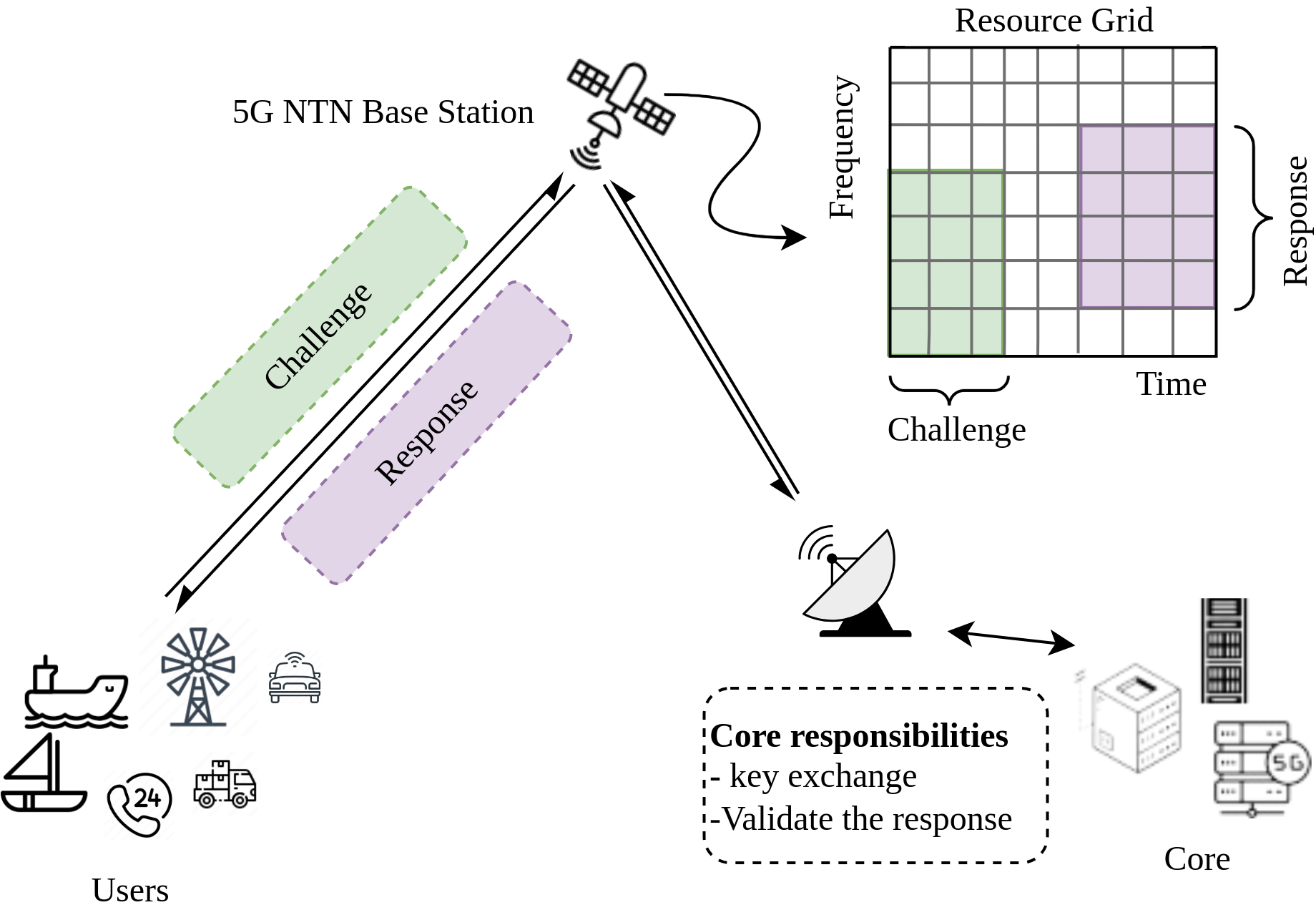}
    \caption{\name can be seamlessly integrated as LEO satellite based base station will have two way communication with UEs. LEO satellites can allocate resources in time and frequency for Distance Bounding.}
    \label{fig:integration_with_5G}
\end{figure}
\section{Conclusion}
GNSS receivers that rely solely on broadcast authentication (e.g., TESLA‐based OSNMA) remain vulnerable to selective delay attacks because an adversary can simply replay the encrypted signals without breaking cryptographic primitives. Verifiable multilateration could solve this problem but demands two way communication with multiple reference nodes, an impractical and costly requirement for global coverage of billions of users. In this paper, we proposed \name that achieves security equivalent to Verifiable Multilateration with far less infrastructure, cost, and communication overhead. We proposed the use of a two way ranging exchange with one LEO satellite and authenticated broadcast GNSS signals, that reduces both hardware cost and communication exchanges. Simultaneously, the use of high altitude GNSS satellites forms a much larger secure geometric bound, significantly increasing the security guarantees.




 \bibliographystyle{plain}
\bibliography{reference}
\newpage

\appendix

\begin{figure*}[t]
  \centering
  \includegraphics[width=\textwidth]{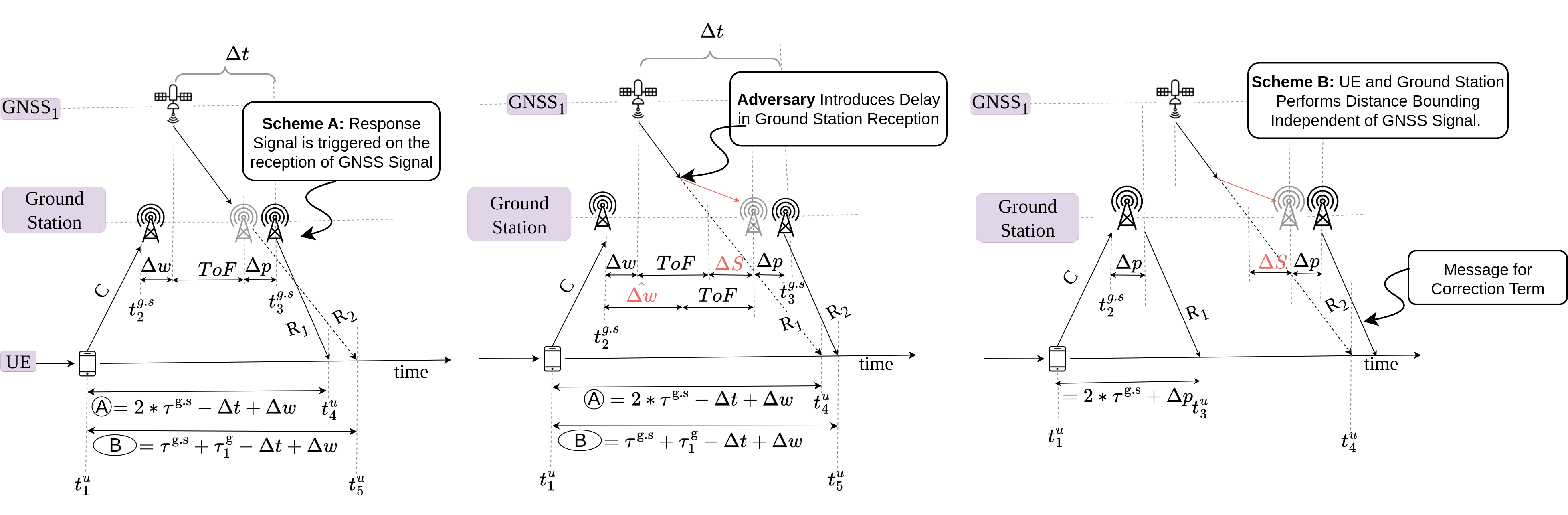}
  \caption{Illustration of clock synchronization vulnerabilities in distance bounding protocols using  reference nodes having insecure clock bias. \textbf{Left:} In Scheme A, the ground station triggers its response based on GNSS signal reception and provides timing offsets to the UE for correction. \textbf{Middle:} An adversary delays the GNSS signal to the ground station, causing it to overestimate the waiting time \(\Delta w\), which leads to a reduction of sum of ToFs. \textbf{Right:} Even in an alternative design where the ground station performs distance bounding independently of GNSS timing, spoofed delays still results in the incorrect estimation of correction term.}
  \label{fig:Appendix}
\end{figure*}

\section{Time Synchronization among Reference Nodes}
The ephemeris data and the associated transmission time in the downlink response signal transmitted by the reference node must be accurate and globally synchronized. Otherwise, spoofed values will cause the UE to obtain an incorrect reference and, in turn, compute an erroneous position. There can be two possible scenarios.
In the first scenario, where the reference node used for distance bounding has both a correct globally synchronized clock and accurate position coordinates, \name is provably secure for position estimation by the end users, as shown in \autoref{sec_analysis}. 
In the second scenario, the reference node has correct knowledge of its position but its clock time is spoofed. This situation can arise, for example, when a ground station acts as the reference node for distance bounding. Since ground stations often derive their timing from GNSS signals, they are susceptible to GNSS time spoofing, making this a realistic threat model. To observe the impact, we consider two protocol variants. 

\subsection{Scheme A}
In Scheme A, the reference node is a ground station that possesses accurate knowledge of its position coordinates but does not have access to a globally synchronized or trusted clock. To facilitate secure distance bounding under this constraint, after receiving challenge signal from the UE, the ground station waits to receive a GNSS downlink signal from a GNSS satellite. Upon receiving this signal at time \(t_2^{\text{g.s}}\), the ground station calculates the geometric ToF \(\tau^{g.s}\) between itself and the GNSS satellite, based on their known positions. Using this, it computes the correction term \(\Delta t\) as the difference in downlink transmission for the correction by UE. 

Additionally, the ground station must communicate waiting time \(\Delta w\) until the reception of GNSS signal and a small processing delay \(\Delta p\), after which it transmits a response signal back to the UE. The UE receives this response and uses the provided timing values; \(\Delta w\), \(\Delta p\), and \(\Delta t\) to reconstruct the correct round trip time and accurately solve for its own position. 

This scheme allows the UE to form a valid ellipsoidal constraint, even though the reference node lacks a correct clock as shown in \autoref{fig:Appendix}

However, if an adversary introduces a spoofing delay of \(\Delta S\) on the GNSS downlink received by the ground station, this causes incorrect position estimation by UE. The spoofed GNSS signal reaches the ground station later than it should, i,e after \( \Delta S \). Because the ground station derives the original GNSS transmission time by subtracting the geometric ToF (computed from satellite and ground station positions) from the spoofed reception time, the estimated transmission time becomes later than its corect value. This causes the ground station to compute an increased waiting time, i.e  \( \hat{\Delta w} =\) \(\Delta w + \Delta S\) and report it to the UE.

Since the UE subtracts the reported waiting time from the total delay, the increased \(\hat{\Delta w}\) leads to a reduced sum of time of flight. 

\subsection{Scheme B}
In Scheme B, the UE and the ground station execute a distance bounding protocol independently of any GNSS signal reception. That is, after receiving the challenge signal from the UE, the ground station immediately processes and sends its response at time \( t_2^{\text{g.s}} \), after a processing delay \( \Delta p \). To estimate the correction term \( \Delta t \), the ground station compares the reception time of a GNSS signal with the geometric ToF \( \tau^{g.s} \) computed from its known position and the GNSS satellite’s position. If an adversary introduces a spoofing delay \( \Delta S \) into the GNSS signal, the signal is received at the ground station later than it should be. As a result, the estimated transmission time is measured later compared to its correct value. When the UE receives this biased \( \Delta t \) and adds it in \autoref{eq:ellipsoid} to remove the negative \( \Delta t \), the UE ends up reducing the total sum in \autoref{eq:ellipsoid}.



%

\end{document}